\documentclass{pasj00}

\newfont{\Sc}{eusm10}

\begin{document}
\SetRunningHead{Uchimoto et al.}{NIR imaging in the $z=3.1$
proto-cluster region}
\Received{}
\Accepted{}

\title{Subaru/MOIRCS Near-Infrared Imaging in the Proto-Cluster Region at $z=3.1$}

\author{ Yuka Katsuno \textsc{Uchimoto},\altaffilmark{1,2,3} 
Ryuji \textsc{Suzuki},\altaffilmark{2,3} 
Chihiro \textsc{Tokoku},\altaffilmark{2.3} 
Takashi \textsc{Ichikawa},\altaffilmark{3} \\
Masahiro \textsc{Konishi},\altaffilmark{2,3}
Tomohiro \textsc{Yoshikawa},\altaffilmark{2,3} 
Koji \textsc{Omata},\altaffilmark{2}
Tetsuo \textsc{Nishimura},\altaffilmark{2} \\
Toru \textsc{Yamada}\altaffilmark{2,3}
Ichi \textsc{Tanaka},\altaffilmark{2,3} 
Masaru \textsc{Kajisawa},\altaffilmark{4} 
Masayuki \textsc{Akiyama},\altaffilmark{2,3} \\
Yuichi \textsc{Matsuda},\altaffilmark{5}
Ryosuke \textsc{Yamauchi},\altaffilmark{6}
and Tomoki \textsc{Hayashino} \altaffilmark{6}} 

\altaffiltext{1}
{Institute of Astronomy, University of Tokyo, 2-21-1 Osawa, Mitaka,
Tokyo, 181-0015}
\email{uchimoto@ioa.s.u-tokyo.ac.jp}
\altaffiltext{2}{Subaru Telescope, National Astronomical Observatory
of Japan,\\ 
650 North A'ohoku Place, Hilo, HI 96720, USA}
\altaffiltext{3}{Astronomical Institute, Tohoku University, Aramaki,
Aoba, Sendai 980-8578}
\altaffiltext{4}
{National Astronomical Observatory, 2-21-1 Osawa, Mitaka, Tokyo, 181-8588}
\altaffiltext{5}
{Department of Astronomy, Kyoto University, Kitashirakawa-Oiwake-cho,
Sakyo-ku, Kyoto 606-8502}
\altaffiltext{6}
{Research Center for Neutrino Science, Graduate School of Science, \\
Tohoku University, Aramaki, Aoba-ku, Sendai, Miyagi, 980-8578}


\maketitle

\begin{abstract}
 We present the results of deep near-infrared imaging observations of the $z=3.1$ proto-cluster region in the SSA22a field taken by MOIRCS mounted on the Subaru Telescope. We observed a 21.7 arcmin$^2$ field to the depths of $J=24.5$, $H=24.3$, and $K=23.9$ (5$\sigma$). We examine the distribution of the $K$-selected galaxies at $z \sim 3$ by using the simple color cut for distant red galaxies (DRGs) as well as the photometric-redshift selection technique. The marginal density excess of DRGs and the photo-z selected objects are found around the two most luminous Ly$\alpha$ blobs (LABs). We investigate the correlation between the $K$-selected objects and the LABs, and find that several galaxies with stellar mass $M_* = 10^{9}-10^{11} M_\odot$ exist in vicinity of LABs, especially around the two most luminous ones. We also find that 7 of the 8 LABs in the field have plausible $K_s$-band counterparts and the sum of the stellar mass possibly associated with LABs correlates with the luminosity and surface brightness of them, which implies that the origin of Ly$\alpha$ emission may be closely correlated with their stellar mass or their previous star formation phenomena.
\end{abstract}

\section{Introduction}
\label{sec:intro}

 The formation history of galaxies has been extensively studied in terms of the evolution of their stellar mass \citep{key-papo01, key-kajisawayamada, key-verma, key-caputi06, key-fontana06, key-arnouts}. However, it is still poorly constrained how galaxies developed their stellar systems in the early universe at $z > 2$. While the observed stellar mass density of the galaxies shows significant and rapid growth along the time at $z > 1$, at least $10$ \% of the present-day stellar mass in the galaxies is observed at $z \sim 3$ \citep{key-caputi06, key-fontana06, key-arnouts}, which seems also true for the massive galaxies with $M_* > 10^{11} M_\odot$ \citep{key-caputi06}. 

Where and how were they formed? The biased galaxy formation models in a universe dominated by cold dark matter suggest that galaxy formation preferentially occurs in regions with relatively high density at the larger scale \citep{key-kauffmann, key-cen, key-benson, key-bower06}. Recent results based on optical wide-field imaging reveal that high-redshift star-forming galaxies indeed show very strong clustering largely biased to the expected underlying mass distribution (e.g., \cite{key-steidel98, key-adelberger, key-ouchi01}, \yearcite{key-ouchi05}), which is also true for near-infrared (NIR)-selected massive galaxies \citep{key-daddi03, key-grazian06, key-quadri07a, key-foucaud, key-ichikawa07}.

 At $z=2-3$, individual large-scale high-density regions of the galaxies have also been discovered \citep{key-steidel98, key-francis,  key-hayashino}. Many of these fields are, however, identified as an excess in the number density of rest-frame UV bright galaxies, such as Ly$\alpha$ emitters (LAEs), Lyman Break Galaxies (LBGs) or their analogues. The question is how much of the stellar mass has already formed and assembled in such high density region of star-forming galaxies at $z\gtsim2$. In other words, we would like to see whether the structure traced by the star formation activity is correlated (or anti-correlated) with the structure traced by the stellar mass. 

 In this paper, we present the results of $J$-, $H$- and $K_s$-band imaging of the $z=3.1$ proto-cluster region in and around the SSA22 field (\cite{key-steidel98}, \yearcite{key-steidel00}; \cite{key-hayashino}) using Multi-Object InfraRed Camera and Spectrograph (MOIRCS) \citep{key-ichikawa06, key-suzuki07} equipped with the 8.2 m Subaru Telescope . The wide field of view of MOIRCS allowed us to observe such proto-cluster regions efficiently. 

 The proto-cluster was first discovered by \citet{key-steidel98} as the excess of LBGs. Later wider-field and deeper narrow-band observations revealed that the overdense region of LBGs is a part of a large-scale high density structure of LAEs, which is extended over $\sim$ 60 Mpc in a comoving scale \citep{key-hayashino, key-matsuda04}. \citet{key-matsuda05} confirmed that the structure has a three-dimensional filamentary appearance. \citet{key-matsuda04} also identified 35 Lyman$\alpha$ Blobs (LABs, extended, bright Ly$\alpha$-emitting clouds), along this structure. The large nebulae typically have Ly$\alpha$ luminosity $>10^{43}$-$10^{44}$ ergs s$^{-1}$ and physical extent 30-150 kpc. The most luminous and extended LABs are the two giant nebulae first discovered by \citet{key-steidel00} (hereafter referred as LAB1 and LAB2, following \cite{key-matsuda04}). Rather than studying the differences in their distribution from only the UV-selected galaxies, observing the star forming regions in NIR, the rest-frame optical wavelengths puts more weight in their stellar mass content, allowing us to study the mass and population of the stars associated with them. The nature of dusty star-forming objects in such high-density structure can also be examined by the NIR data.

The field observed with MOIRCS (hereafter referred as SSA22-M1) corresponds to the southern part of SSA22a \citep{key-steidel00} and is located within the field observed by \citet{key-matsuda04}.
 In SSA22-M1, there are 8 LABs \citep{key-matsuda04}, 16 LAEs \citep{key-hayashino}, and 17 LBGs  \citep{key-steidel98, key-steidel03b}, of which 11 are located at $z=3.1$. The sample selection for LAEs is overlapped with that for LABs. While LAEs are selected by the aperture photometry of the detected sources \citep{key-hayashino}, LABs are selected by their isophotal area on the narrow-band image reduced with the optimum sky subtraction \citep{key-matsuda04}. Some LABs are associated with LAEs if they have enough compact high-surface brightness cores. On the other hand, Hayashino et al. (2004) excluded the LAEs detected in LAB1 and LAB2 from their sample as they are clearly un-isolated parts of diffuse Ly$\alpha$ nebulae. Two of the 16 LAEs in SSA22-M1 are identified with LABs (LAB16 and LAB31, referred by \cite{key-matsuda04}). 

 We describe the observation and the data analysis method in section 2. The number density and the sky distribution of distant red galaxies (DRGs, \cite{key-franx}), the result of photometric redshift analysis of the $K$-selected sources, and the photometric properties of LABs, LBGs, and LAEs are shown in section 3. In section 4 we discuss our results in terms of the mass assembly history in the proto-cluster. We use the cosmological parameter values $\Omega _{\rm M}=0.3$, $\Omega _{\Lambda}=0.7$, and $H_0$=70 km s$^{-1}$ Mpc $^{-1}$ throughout this paper. All the magnitude values are in {\it AB} system \citep{key-oke, key-fukugita} unless explicitly noted.

\section{Observation and Data Analysis}
\label{sect:observation}

 $J$-, $H$-, and $K_s$-band images of the field centered at ($\alpha, \delta$)$_{2000}$= (22h 17m 33.9s, $+00' 12'' 14''$) were obtained using MOIRCS mounted on the Subaru Telescope on June 15, 16 and August 14 in 2005, and July 23 in 2006 (UT). MOIRCS uses two 2048$\times$2048 HAWAII-2 arrays with the pixel scale $0''.117$ pixel$^{-1}$ and the field of view $4'\times 7'$. 

 The images are reduced in a standard manner with the MCSRED software packages (I. Tanaka et al. 2007, in preparation). For the $K_s$-band flat-fielding, we use on/off sets of domeflat images to avoid the thermal emission seen at the edge of the frames due to the foreground telescope structure. The median-sky for each frame is  subtracted before the frames are stacked. If notable diffuse residual patterns remain after the median-sky subtraction, we further subtracted them by using the IRAF {\sf imsurfit} command. We also subtract fringe patterns in each frame, if they appear, using the templates made from the images taken in the same run.

 The total exposure times of the final images are 3,240 s, 3,600 s, 2,235 s for $J$, $H$, $K_s$, respectively.  The effective area of the image is 21.7 arcmin$^2$. The stellar image sizes are $0''.51$, $0''.47$, $0''.44$ for $J$, $H$, $K_s$, respectively. The limiting magnitudes are $J= 24.5$, $H= 24.3$, and $K = 23.9$ with 5 $\sigma$ in a $1.1''$ diameter aperture. We calibrated our NIR data to the UKIRT photometric system \citep{key-tokunaga}. We convert between the Vega system to the AB system using $J=J_{\rm Vega}+0.95$, $H=H_{\rm Vega}+1.39$, $K=K_{\rm Vega}+1.85$. The summary of the observations is shown in Table \ref{tab1}.  

 For the detection and photometry, we use SExtractor (version 2.3) \citep{key-bertin}. The objects which have more than 16 connected pixels with 1.5 $\sigma$ above the background noise are selected. The $J-K$ color is measured in a $1''.1$ diameter aperture. The SExtractor MAG AUTO value is adopted as the $K$-band pseudo total magnitude. The $K_s$-band image is smoothed to be matched with the PSF in $J$-band for $J-K$ color measurement. The completeness in $K_s$-band is 95$\%$ at $K=23.5$ and 50 $\%$ at $K=23.9$ for a point source.

 We also use the optical $BVRi'z'$ bands data obtained by
 \citet{key-hayashino}  and $U_n$-band data
 \citep{key-steidel03a}\footnote{ftp://ftp.astro.caltech.edu/pub/ccs/lbgsurvey/} to estimate the photometric redshift. As the FWHM of the  $U_n$-band  image is $1''.3$, we smoothed the $BVRi'z'JHK_s$-band images to be matched and measured the colors in a $2''.6$ aperture. 
Narrowband (NB497; 4977 \AA, FWHM 77 \AA) images used in section \ref{sec:lab} are
taken in \citet{key-matsuda04}.

We corrected our data for the Galactic extinction adopting the average value at the field,  $E(B-V) = 0.08$ \citep{key-schlegel}.

\begin{table}
\begin{center}
\caption{Summary of the NIR observations\label{tab1}}
\begin{tabular}{cccc}
\hline
\hline
\noalign {\vspace{0.15cm}}
Filter & Integration & FWHM & Limiting Mag.\\
& s & arcsec & mag \footnotemark[*]  \\
\noalign {\vspace{0.15cm}}
\hline
\noalign {\vspace{0.15cm}}
$J$ & 3,240  & 0.51 & 24.5 \\
$H$ & 3,600  & 0.47 & 24.3 \\
$K_s$ & 2,235  & 0.44 & 23.9\\
\noalign {\vspace{0.15cm}}
\hline
  \multicolumn{4}{@{}l@{}}{\hbox to 0pt{\parbox{180mm}{\footnotesize
\footnotemark[*] The 5$\sigma$ limit for a point source in a $1.1''$ diameter aperture.
}\hss}}
\end{tabular}
\end{center}
\end{table}

\section{Results}

 In this section, we present the observed properties of the $K$-selected objects in SSA22-M1. Since we are studying the galaxies at $z=3.1$, we need to apply some photometric selections to suppress the contamination of foreground/background objects. While full photometric redshift analysis is useful, it may suffer from misidentification especially for $K$-band faint objects because the optical data used here may not be deep enough for them to achieve high photometric accuracy. Simple color-cut method such as selecting DRGs with $(J-K)_{\rm Vega} > 2.3$ \citep{key-franx} works complementary, so that we can more completely pick-up objects possibly located at $z=3.1$, if they exist. We therefore first investigate the distribution of DRGs and then that of the photometric redshift (photo-z) selected sources to compare with the UV-selected objects in the field.

\subsection{The Selection and Number Density of DRGs}
\label{sec:drg}

 We first focus on the distribution of DRGs in our field. DRGs are supposed to be evolved, relatively massive galaxies ($M_* \sim 10^{10}-10^{11} M_\odot$) at $2 \ltsim z \ltsim 4$ \citep{key-franx, key-forsterschreiber, key-reddy, key-ichikawa07} or dusty galaxies at $z \gtsim 1$ \citep{key-webb06, key-conselice, key-quadri07b, key-lane}. 
In the analysis of the deep multicolor data obtained at GOODS-N, \citet{key-kajisawa06} found that 83\% of the DRGs with $22 < K < 23.5$ have the photometric redshift $z > 2$, while more than a half of the DRGs with $K < 22$ have $z < 2$. Thus the faint DRGs are supposed to be dominated by the galaxies at high redshift. 
As \citet{key-marchesini} suggested that the global stellar mass at $2\ltsim z \ltsim 3.5$ appears to be dominated by DRGs, the number of DRGs might be a good indicator of massive galaxies at $z\gtsim2$. In fact, \citet{key-labbe05} showed that DRGs dominate the high mass end of the mass function at high redshift. 

 We use the same criteria as in \citet{key-kajisawa06}, which corresponds to $(J-K)_{AB}>1.4$ in our photometric system. 
The red $J-K$ color of DRGs is due to the Balmer or 4000\AA\ break of galaxies at $z \gtsim 2$, or due to the dust extinction \citep{key-franx, key-reddy, key-forsterschreiber}.  The color $(J-K)_{\rm AB}>1.4$ corresponds to that of a galaxy at $z=3.1$ with the stellar population age older than 200 Myr if no extinction and the solar metallicity are assumed \citep{key-bruzual}.  
\citet{key-forsterschreiber} reported the median age and stellar mass for DRGs with $K<25.6$ in HDF-S are 1.7 Gyr and $M_*=0.8 \times 10^{11} M_{\odot}$. 
The $K$-band selected DRGs are supposed to be relatively more massive, more evolved objects than ultra-violet(UV)-selected galaxies like LBGs at the same epoch.

\begin{figure}
\begin{center}
\FigureFile(80mm,80mm){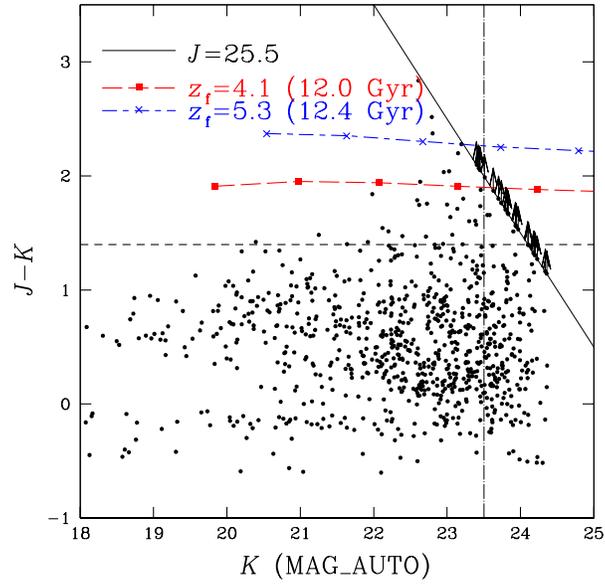}
\end{center}
\caption{$K$ vs. $J-K$ 
color-magnitude diagram of the $K$-selected objects. The solid line indicates the detection limit of 2$\sigma$ above the sky fluctuation in $J$-band. The vertical dot-dashed line shows the $K$-band completeness limit of 95 \%. The horizontal dashed line shows $J-K=1.4$. The expected color-magnitude relations of cluster galaxies calculated by \citet{key-kodama97} for Coma cluster model with the formation redshift at $z_f=4.1$  and $z_f=5.3$ are plotted with the squares and the crosses, respectively. The points indicate the location of model galaxies with the rest-frame $V$-band luminosity from $M_V=-17$ to $M_V=-22$ at $z=0$. The objects with $J-K \ltsim 0$ are expected to be galactic stars.}
\label{fig_colmag}
\end{figure}

\begin{figure}
\begin{center}
\FigureFile(80mm,80mm){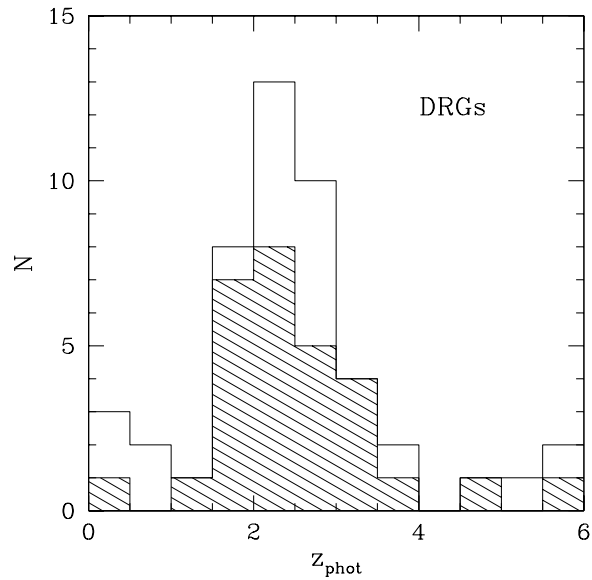}
\end{center}
\caption{
Histogram of the estimated photometric redshift of DRGs.
The hashed histogram is DRGs with $K<23.5$.
}
\label{fig_photozP50}
\end{figure}

 In Fig. \ref{fig_colmag}, we show the $K$ versus $J-K$ color-magnitude diagram of the $K$-selected objects. The objects with $J-K \ltsim 0$ are likely to be galactic stars. The color of the sequence coincides with that seen in the other field \citep{key-kajisawa06}, which suggests that our photometric calibration is robust. The 2$\sigma$ limiting magnitude in $J$-band is indicated by the solid line in Fig. \ref{fig_colmag}. The $J$-band image is deep enough to select DRGs down to $K=24.1$. 
The distribution of the photometric redshifts for DRGs is shown in Fig. \ref{fig_photozP50}. The detailed photo-z analysis is described in section \ref{sec:photoz}. 
We find that 62 \% of DRGs are classified as $2 < z_{\rm photo} < 4$. 

\begin{figure}
\begin{center}
\FigureFile(80mm,80mm){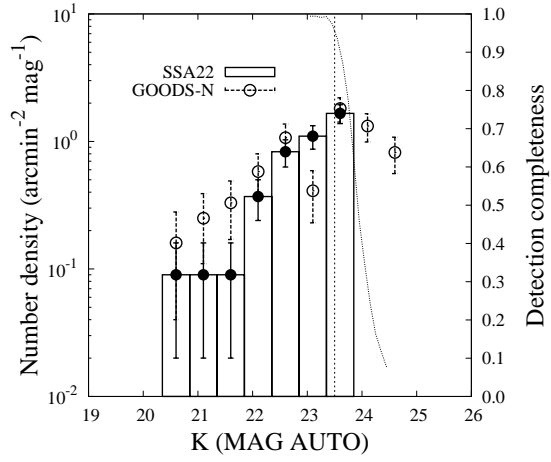}
\end{center}
\caption{
Differential number counts for DRGs in our observed field (filled circles) in comparison with those for DRGs in the GOODS-N field (open circles) \citep{key-kajisawa06}. The vertical dotted line shows the $K$-band completeness limit of 95 \%. The detection completeness in $K$-band is also shown with the smaller dotted line.
}
\label{fig_drg_density}
\end{figure}

 The cumulative number of DRGs in our observed field is 28 and the surface density is 1.29 arcmin$^{-2}$ at $K<23.35$ (or $K_{\rm Vega} < 21.5$), which is comparable with that in GOODS-N by \citet{key-kajisawa06}, 1.36 $\pm$ 0.26 arcmin$^{-2}$ to the same depth. Fig. \ref{fig_drg_density} compares the differential number counts for DRGs in our observed field with those in GOODS-N. Our sample is dominated by DRGs with $22<K<23.5$, which implies most of them should be at high redshift. The completeness fraction and the 95\% limit, $K=23.5$, are also marked. We do not see any notable excess in the number counts in SSA22-M1 over the magnitude range.

We note that  there are nine hyper extremely red objects (HEROs, \cite{key-totani}) with $(J-K)_{AB}>2.1$ in SSA22-M1 at $K<23.5$. The surface density of HEROs at $K<23.5$ is $0.41 \pm 0.14$ arcmin$^{-1}$ which is consistent with those in SDF, HDF-S, and GOODS-N within the error \citep{key-maihara, key-saracco, key-kajisawa06}.

\begin{figure}
\begin{center}
\FigureFile(80mm,80mm){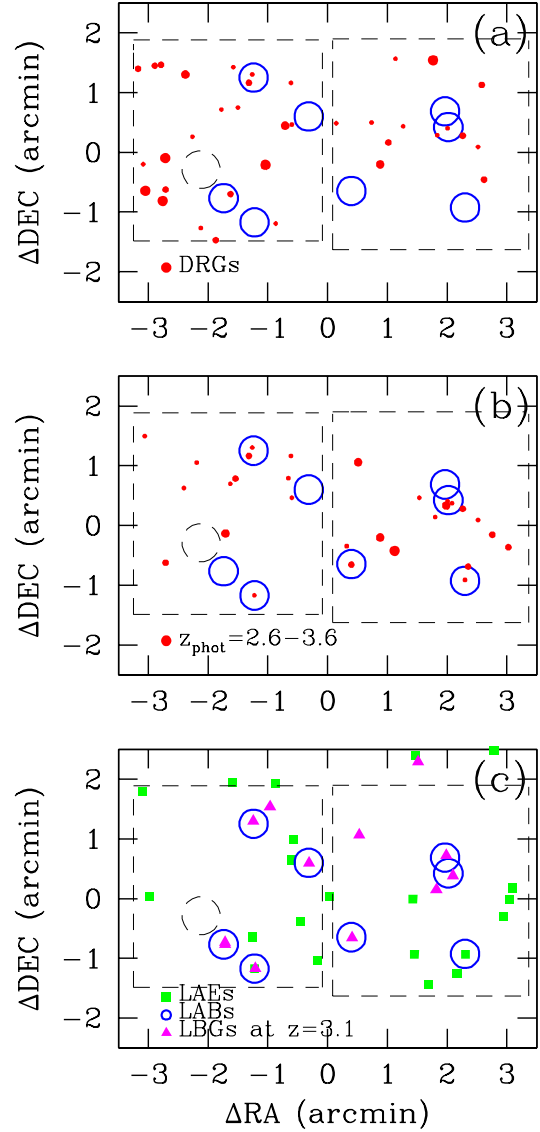}
\end{center}
\caption{
(a) Sky distribution of DRGs and LABs. DRGs are indicated with filled circles (red). LABs  are indicated with large open circles (blue). Dotted squares indicate the observed regions with MOIRCS. The dotted large circle at the east side is the region excluded from the analysis due to a bright star. (b) Sky distribution of $z_{\rm phot}=2.6-3.6$ objects with $K<23.5$ and LABs. The objects are indicated with filled circles (red). (c) Sky distribution of LABs, LAEs and LBGs. LAEs (square, green) from \citet{key-hayashino} and LABs (large open circle, blue) from \citet{key-matsuda04} are shown. LBGs at $z=3.1$ from Steidel et al. (2003) are indicated with triangles (magenta).
}
\label{fig_geo}
\end{figure}

\begin{figure*}
\begin{center}
\FigureFile(150mm,80mm){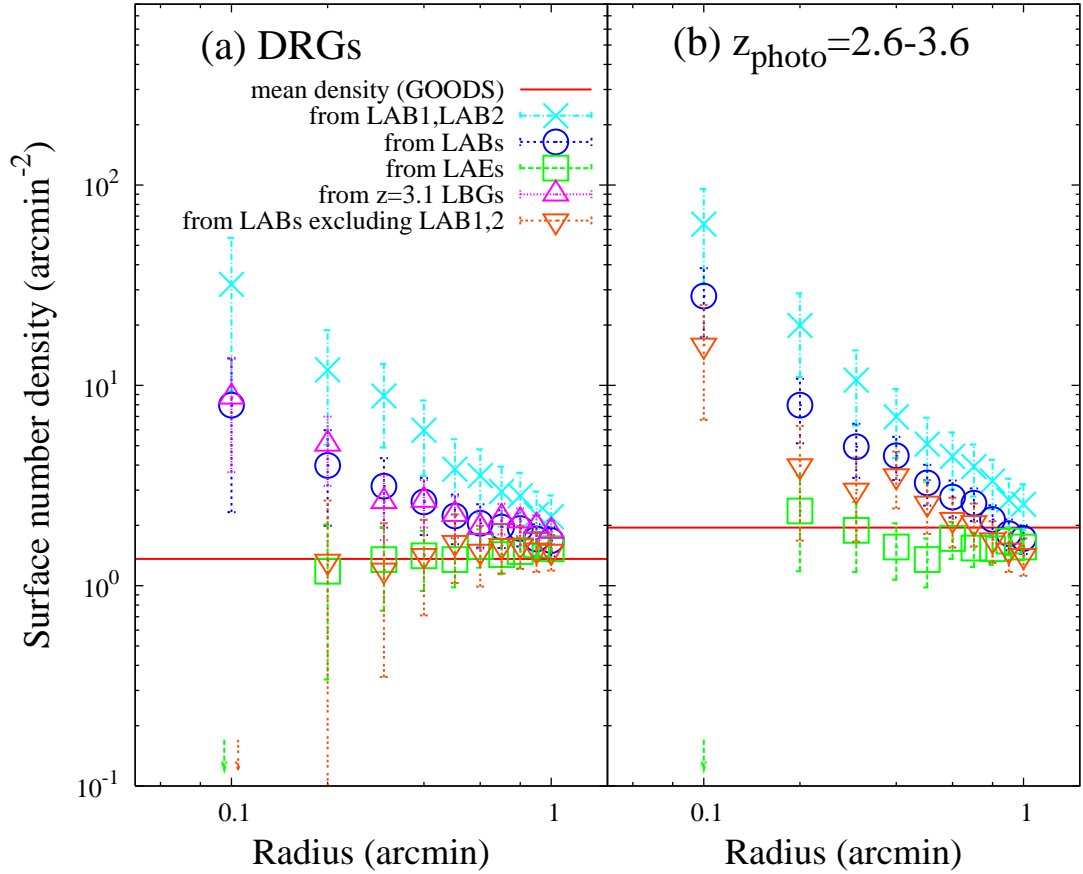}
\end{center}
\caption{
(a) The mean surface number density of DRGs around optically selected galaxies. The horizontal axis is radial distance from LABs, LAEs or LBGs. The surface densities of DRGs from LAB1 and LAB2, LABs, LAEs, LBGs, and LABs excluding LAB1 and LAB2 are shown as crosses (cyan), open circles (blue), open squares (green), open triangles (magenta), and inverted triangles (orange), respectively. The mean density of DRGs in GOODS-N \citep{key-kajisawa06} is also indicated with a solid line. The arrows indicate zero at $r=0.1$ arcmin.
(b) The mean surface number density of $z_{\rm phot}=2.6-3.6$ objects with $K<23.5$ around optically selected galaxies. The mean density of $z_{\rm phot}=2.6-3.6$ objects in GOODS-N \citep{key-ichikawa06} is also indicated with a solid line.
}
\label{fig_box}
\end{figure*}

 The sky distribution of DRGs is shown in Fig. \ref{fig_geo}a, and LABs \citep{key-matsuda04}, LAEs \citep{key-hayashino}, and LBGs \citep{key-steidel03b} are plotted in Fig. \ref{fig_geo}c.  We then investigated the spatial correlation between DRGs and other UV-selected objects in the field. Fig. \ref{fig_box}a shows the mean surface number density of DRGs as a function of a distance from LABs, LAEs, and LBGs. 

First, we note that the distribution of the DRGs is weakly correlated with the sample of LABs as well as LBGs. Since the LBGs are associated with 7 of 8 LABs, it is not surprising that the similar correlations with DRGs are seen for these two populations. We also plotted the profile from the two distinctive giant LABs, namely LAB1 and LAB2 and found that the DRG density shows the notable excess within 1.0 arcmin (1.9 Mpc at $z=3.1$ in comoving scale) around them. In fact, the correlation between DRGs and LABs is found to be dominated by the two objects; if we remove them from the sample, the correlation seen in Fig. \ref{fig_box}a diminishes significantly. 

As mentioned in the beginning of this section,
the redshift range of the DRG criteria is broad and there may be a chance
projection of interlopers at intermediate redshift.
We made a simple test to see if the density excess of DRGs around LAB1 and LAB2 is significant by evaluating the probability of finding similar excess around the eight randomly placed objects. We find the similar profile in 592 among the 10000 trials, which gives the probability of $\approx 6\%$ in random distribution of LAB to the observed distributions DRGs. Thus the association of LAB1/LAB2 with DRGs is marginally significant ($\approx 2\sigma$). We investigate the closer look for the $K$-selected objects near the eight LABs in the field in section \ref{sec:lab}.

We also find that there is no significant positive correlation between DRGs and LAEs, while a hint of anti-correlation is seen at $\sim 6''$ separation.  It should be noted that there is such anti-correlation among the massive DRGs and less massive LAEs.

\subsection{Photometric Redshift of the $K$-selected Sources}
\label{sec:photoz}

 Using the $U_nBVRi'z'JHK$ photometric data, we estimated the photometric redshift of all the $K$-selected sources using the HyperZ code \citep{key-bolzonella}. The photometric redshift method performs the SED fitting with the redshift, spectral type, age, and dust extinction as free parameters. The best fit SED with its redshift is determined from the minimum $\chi^2$ value. The template spectrum we use are derived from GALAXEV, which is the library of evolutionary stellar population synthesis models by \citet{key-bruzual}. 

\begin{figure}
\begin{center}
\FigureFile(80mm,80mm){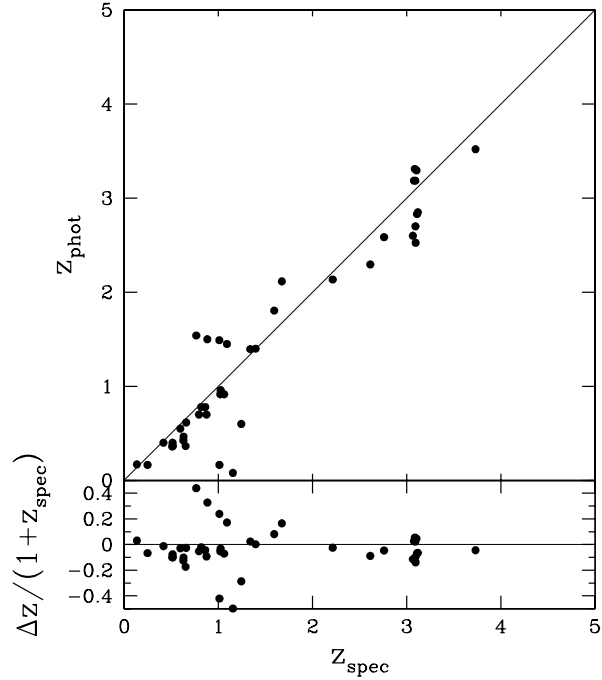}
\end{center}
\caption{
Spectroscopic and photometric redshifts of galaxies 
for the $K$-selected objects in the SSA22-M1 field.
The relative errors $(z_{\rm photo} - z_{\rm sp})/(1+z_{\rm sp})$ are 
indicated in the bottom panel.
}
\label{fig_speczphotoz}
\end{figure}

At $z \sim 3$, the uncertainty of the photo-z is typically $\Delta z \sim 0.5$ \citep{key-kajisawa05, key-reddy, key-cirasuolo, key-ichikawa07}.
Fig. \ref{fig_speczphotoz} shows the resulting photometric redshifts versus spectroscopic redshifts obtained by previous studies (\cite{key-songaila, key-cowie, key-steidel99, key-steidel03b, key-abraham, key-swinbank, key-doherty}). The relative errors $(z_{\rm photo} - z_{\rm sp})/(1+z_{\rm sp})$ in the redshift estimation are also shown in the bottom panel of the figure. 
The mean of relative errors for LBGs at $z\sim 3$ is $-0.04$ and the standard deviation is 0.08 (also see Table \ref{tab3} in section 3.3).
We therefore pick up the objects with $2.6 < z_{\rm phot} < 3.6$ as the candidate galaxies at $z=3.1$. 
We found 29 objects of which 9 are classified as  DRGs and 6 are classified as LBGs. 

 Fig. \ref{fig_geo}b shows a sky plot of the $K$-selected sources, similar to Fig. \ref{fig_geo}a, but for those with $2.6 < z_{\rm phot} < 3.6$ and $K<23.5$. Four LBGs associated with LABs are not appeared since they are fainter than the limit in $K$-band. Cross correlation with LABs, LAEs, and LBGs for this sample is also shown in Fig. \ref{fig_box}b. We find that the association between these $K$-selected sources at $z=3.1 \pm 0.5$ and LABs, especially LAB1 and LAB2, seems significant. In contrast to DRGs, LABs excluding the two giant blobs show the association with the photo-z selected objects.

\subsection{NIR Properties of the LABs}
\label{sec:lab}

\begin{figure*}
\begin{center}
\FigureFile(150mm,80mm){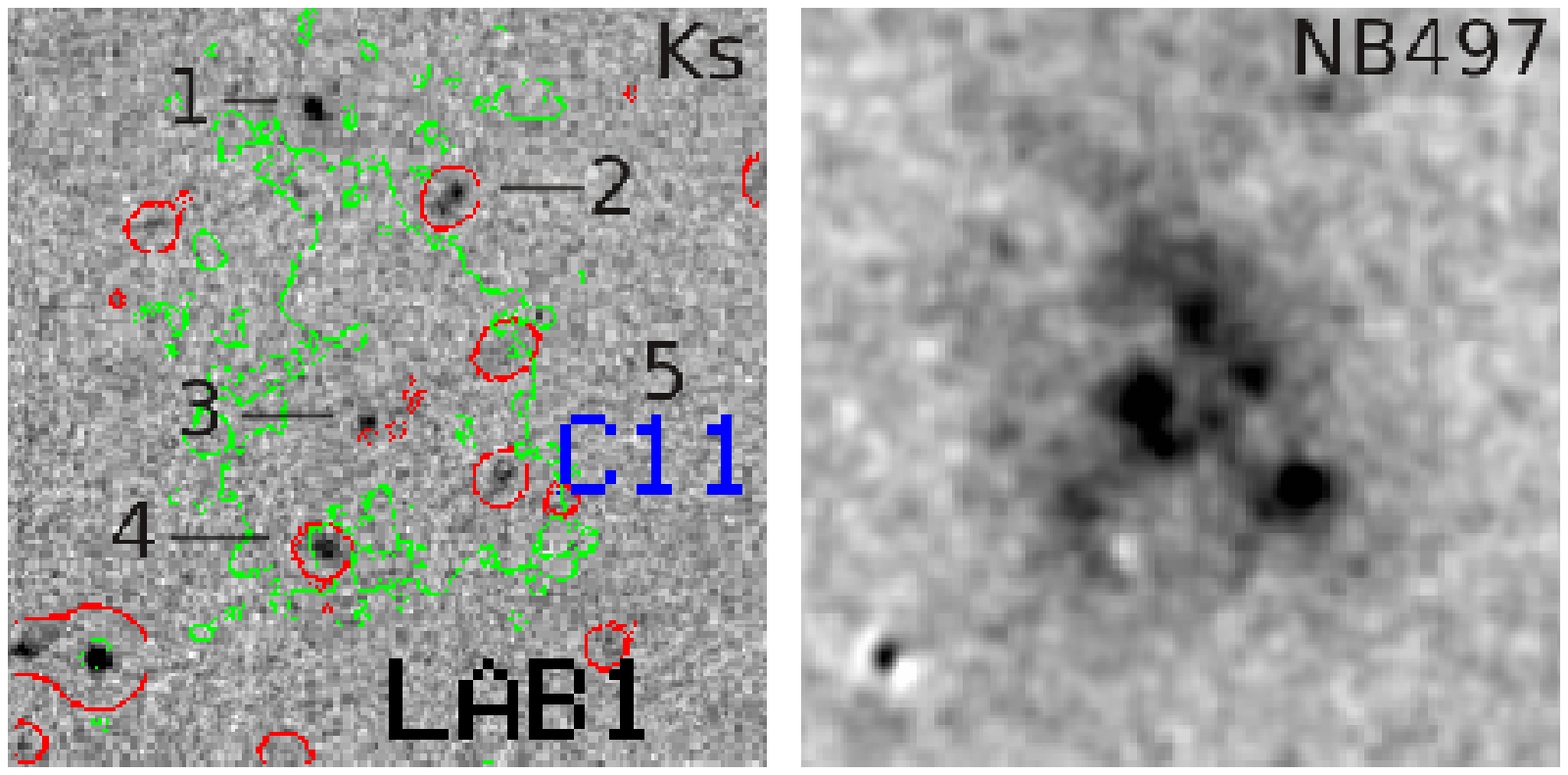}
\FigureFile(150mm,80mm){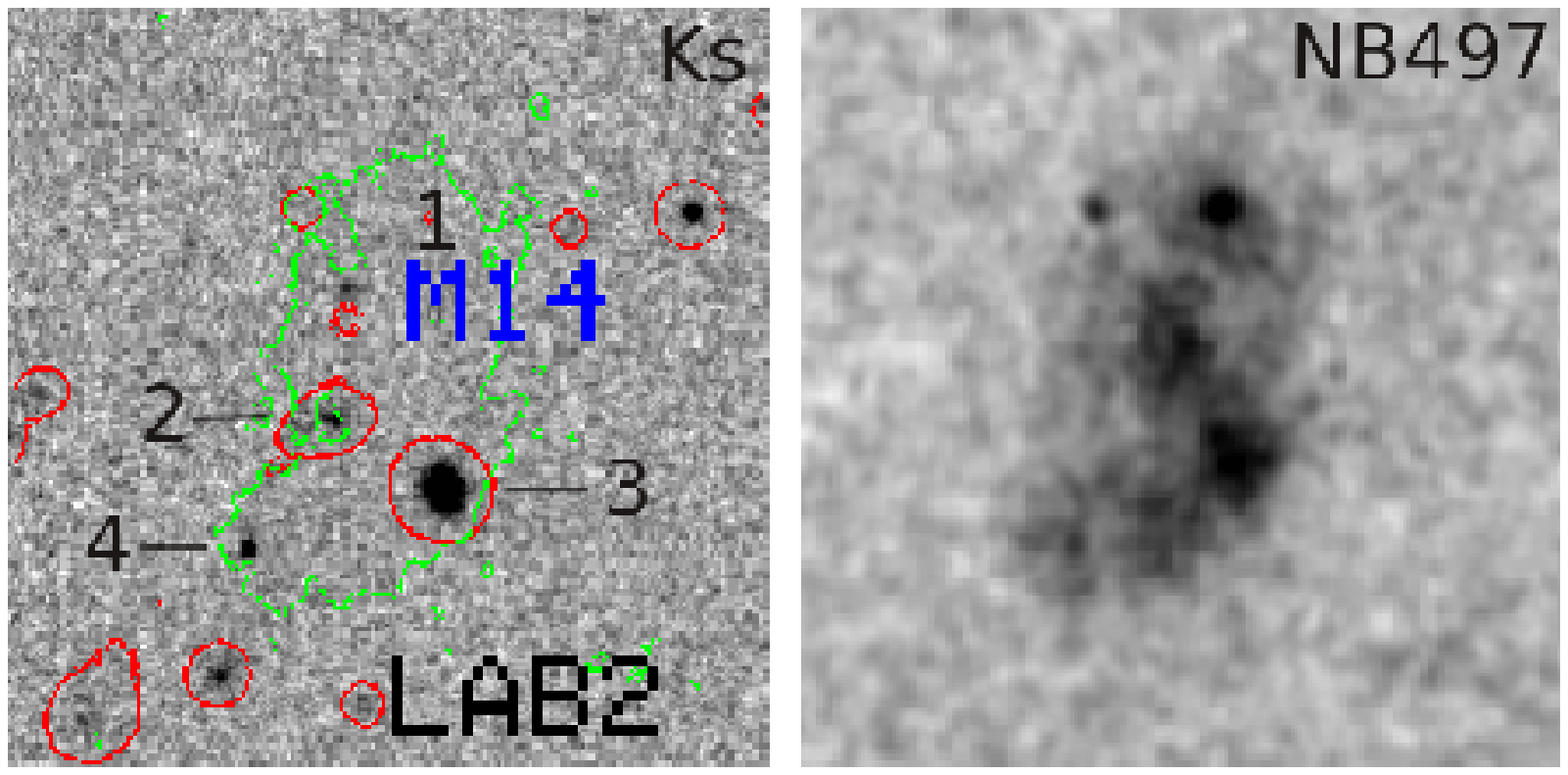}
\end{center}
\caption{
$K_s$-band and narrowband (NB497; 4977 \AA, FWHM 77 \AA) images around LAB1 and LAB2 at $z=3.1$ \citep{key-matsuda04}. The size of each panel is $25''$, which corresponds to $\sim 190$ kpc at $z=3.1$. Each image is centered on a LAB from \citet{key-matsuda04}. The green and red contours are the isophotal levels of NB497 and $R$-band images, respectively.
}
\label{fig_lab1lab2}
\end{figure*}

\begin{figure*}
\begin{center}
\FigureFile(150mm,150mm){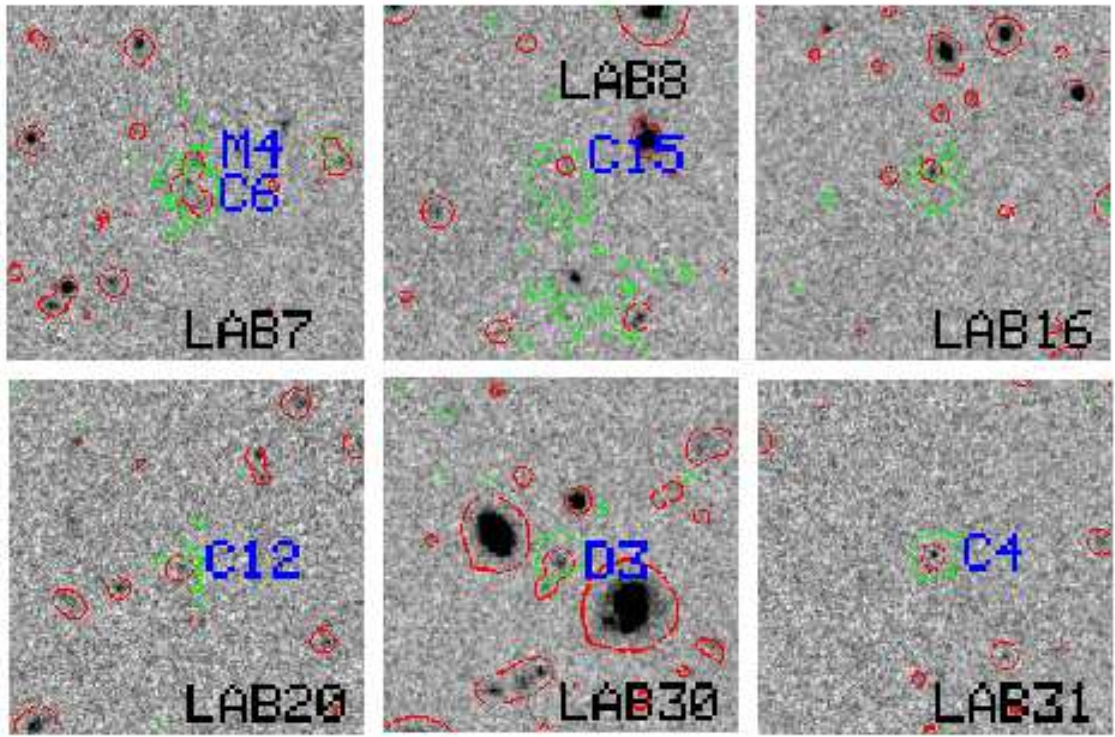}
\end{center}
\caption{
$K_s$-band images around LABs at $z=3.1$ \citep{key-matsuda04}. The size of each panel is $25''$, which corresponds to $\sim 190$ kpc at $z=3.1$. Each image is centered on a LAB from \citet{key-matsuda04}. The green and red contours are the isophotal levels of NB497 and $R$-band images, respectively.
}
\label{fig_lab7to31}
\end{figure*}

 In this section, we investigate the NIR counterparts of the individual LABs. We find that the 88\% of the LABs have plausible $K_s$-band candidates. $K_s$-band images in $25''$-side box for 8 LABs are shown in Fig. \ref{fig_lab1lab2} and Fig. \ref{fig_lab7to31}. The green lines indicate the $NB497$-band detection isophoto contours \citep{key-matsuda04} and $R$-band sources are indicated with the red contours in the figures. LBGs in the field \citep{key-shapley,key-steidel03b} are also marked. Of 11 LBGs at $z=3.1$, 8 are associated with LABs. Of these, SSA22a-C11 (Steidel et al. 2003, associated with LAB1), M14 (LAB2), M4 (LAB7), C12 (LAB20), D3 (LAB30), and C4 (LAB31) are detected in the $K_s$-band, while C6 (LAB7) and C15 (LAB8) are not. 

Photometric properties of $K$-selected objects around LABs are
summarized in Table \ref{tab2} and the evaluated photometric redshift
and physical properties of the candidates of $z=3.1$ galaxies are
presented in Table \ref{tab3}. 
The stellar age, dust extinction, absolute magnitude, and stellar mass are derived from the SED fitting with GALAXEV \citep{key-bruzual} by assuming $z=3.1$. 
We assume the \citet{key-chabrier} initial mass function (IMF), the \citet{key-calzetti} extinction law, and a solar metallicity. 
The characteristic time scale is set at $\tau=0, 1, 9$ Gyr.
The star formation timescale, the color excess, and the age are varied 
as free parameters and 
the best fit SED is determined from the minimum reduced $\chi^2$ value. 
The mass-to-light ratio $(M/L)_V$ and the total absolute magnitude ($L_V$) derived from GALAXEV are used for calculating the stellar mass. The errors of the stellar mass indicate the confidence level of 68 \%.

 For LAB1, there are five $K$-selected sources within the Ly$\alpha$ nebula. 
LAB1-\#1 and \#2 are likely to be foreground objects as their photometric redshifts are $z_{\rm phot}<1$.
\citet{key-steidel00} already reported that there is the $K_s$-band counterpart candidate with extremely red $R-K$ color near the center of the Ly$\alpha$ nebula (LAB1-\#3 in Fig. \ref{fig_lab1lab2}). We find that LAB1-\#3 is classified as a DRG and
LAB1-\#3 and \#4 have $2.6 < z_{\rm phot} < 3.6$. While their spectroscopic redshifts are still unknown, it is very likely that \#4 associates the LAB as it is located right at the hole of Ly$\alpha$ nebulae (the right panel in Fig. \ref{fig_lab1lab2}); it may be an object which absorbs the Ly$\alpha$ emission at the redshift \citep{key-matsuda07}. \#3 and \#4 are also detected in Spitzer IRAC images in 3.6-8.0 $\mu$m (\cite{key-geach07}; T. Webb et al. 2007, in preparation). The $K_s$-band counterpart of SSA22a-C11 at $z=3.109$ is also detected (\#5). 

 For LAB2, four $K$-selected sources are found. LAB2-\#1 is classified as a DRG and it can be the counterpart of SSA22a-M14 ($z=3.091$) although it is located at $0''.9$ north from the $R$-band position as mentioned in \citet{key-ohyama}.   Probably, we see a mature or dusty part of the object in the $K_s$-band image.  
LAB2-\#2 is likely to be foreground objects as the photometric redshift is $z_{\rm phot}<1$.
LAB2-\#4 is also classified as a DRG with extremely red color, $R-K=4.32$. 

 In LAB7, \citet{key-steidel03b} and \citet{key-shapley} sampled two LBGs, SSA22a-M4 ($z=3.093$) and C6 ($z=3.092$), whereas they are separated as three objects in $BVRi'z'$-band images of \citet{key-hayashino}. Ly$\alpha$ emission entirely covers the three objects. From the coordinate we may identify the north component of this lump as LBG SSA22a-M4, which is detected in our $K_s$-band image. On the other hand, other two components are not detected either in the $J$ nor in the $K_s$-band image.

 LAB8 is located at $15''$ north of LAB1 and may form one large object \citep{key-matsuda04}. There is a LBG SSA22a-C15 ($z=3.094$), which is not detected in either $J$ nor in $K_s$-band.

 LAB16 is not associated with a LBG. We find a counterpart candidate in $K_s$-band whose $R-K$ color is slightly redder than those of typical LBGs at $z=3.1$. The photometric redshift of the object is $z_{\rm phot}=2.7$.
 
 LAB20, 30, and 31 are associated with the LBGs SSA22a-C12 ($z=3.118$), D3 ($z=3.086$), and C4 ($z=3.076$), respectively. The $K_s$-band counterparts are detected for all of them.

 Of the 16 LAEs, only two are detected in $K_s$-band and they are both associated with the LABs (LAB16 and LAB31, see above). From the upper limit of their $K$-band flux, for the rest 14 LAEs, we estimate that their stellar masses are lower than $\sim 2 \times 10^{9}$ M$_\odot$, assuming the spectrum of very young (10$^6$-yr old) objects.


\section{Discussion}

 We here discuss whether massive and/or mature galaxies have already formed in the cluster or proto-cluster which was characterized by the overdensity of star-forming galaxies. 

There is the evidence that at least some relatively massive young galaxies ($10^{10}\sim10^{11}$ \MO) have been formed in the proto-cluster. Fig. \ref{fig_box}b shows that the $K$-selected photo-z sample exhibits an excess within a radius of 0.1-0.7 arcminutes from the LABs. This indeed suggests that LABs are related to the formation of massive galaxies \citep{key-matsuda06}. We also found that DRG density shows a notable excess around LAB1 and LAB2. How unique are LAB1 and LAB2 among the 35 LABs in the same structure? They are the largest and the most luminous in Ly$\alpha$ emission. Moreover, \citet{key-matsuda05} suggested that LAB1 and LAB2 are located at the intersection of the filamentary structure of LAEs. Our results support the picture that LAB1 and LAB2 sit near the center of the proto-cluster, where the growth of the galaxy structure occurs most preferentially and we are witnessing a stage where a significant amount of the stars is being formed in such regions. 

 We also investigated the stellar mass of the $K$-selected objects, which are expected to be associated with LABs, by assuming they are located at $z=3.1$ as described in section \ref{sec:photoz}. The result is also tabulated in Table \ref{tab3}. Although the uncertainty of the stellar mass derived from the SED fitting is large, it is shown that the stellar mass of the LAB counterparts ranges from $ 4 \times 10^{9}$ M$_\odot$ to $ 1 \times 10^{11}$ M$_\odot$. 

\begin{figure*}
\begin{center}
\FigureFile(150mm,150mm){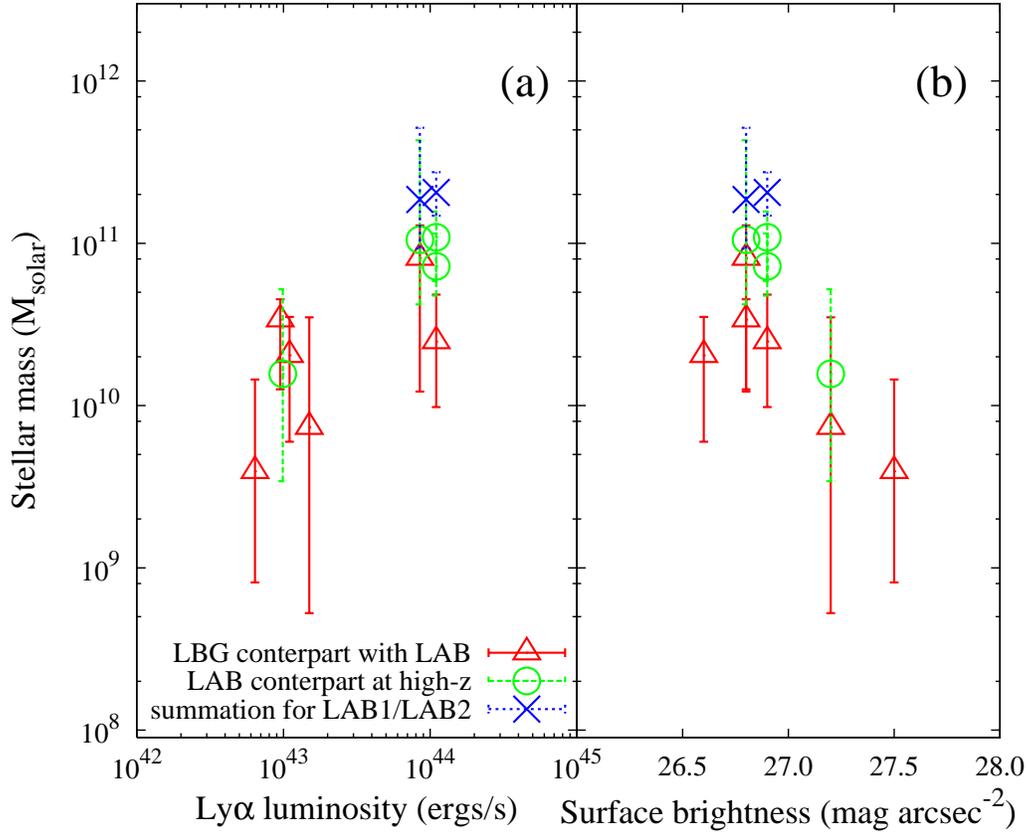}
\end{center}
\caption{ 
(a) Ly$\alpha$ luminosity vs. the stellar mass of the NIR-counterpart candidates. (b) Surface brightness of Ly$\alpha$ vs. the stellar mass of the NIR-counterpart candidates (see text). The redshift of NIR objects is assumed to be $z=3.1$. LBG counterparts associated with LABs are indicated with open triangles (red) and $K$-selected objects expected to be associated with LAB, but with no LBGs, are indicated with open circles (green). Since LAB1 and LAB2 include several NIR counterparts, the total stellar masses in each LAB are indicated with blue crosses.
} 
\label{fig_lya}
\end{figure*}

 Fig. \ref{fig_lya}a shows the relation between Ly$\alpha$ luminosity and the integrated stellar mass of the $K$-selected objects which are associated with each LAB, i.e., the objects in Table \ref{tab3}. Surface brightness of Ly$\alpha$ vs. the stellar mass is also shown in Fig. \ref{fig_lya}b.  These figures suggest that the more massive galaxies are seen in the brighter LABs which have the higher surface brightness in Ly$\alpha$. This result implies the origin of the Ly$\alpha$ emission of LABs may be related with their previous star-formation history.

 Since the local or intermediate-redshift clusters are dominated by the passively-evolving old galaxies, it is worth constraining how many such passive galaxies formed at the higher redshift exist, if any, in the proto-cluster. In Fig. \ref{fig_colmag} we plotted the expected color-magnitude relation for the model galaxies with $M_V=-17$ to $M_V=-22$ at $z=0$ of the Coma metallicity-sequence model \citep{key-kodama97}. It is clearly seen that few objects have color and magnitude for such passive galaxies formed at $z_F > 4$. This result seems to be consistent with that in \citet{key-kodama2007}, which reported that the bright end of the red sequence in proto-clusters around radio galaxies appeared at $z\sim2$, whereas it was not seen at $z\sim3$.

\citet{key-steidel00} reported that the volume density of LBGs at $z\sim3.1$ in the SSA22a/b fields, with the comoving volume of $2.7\times 10^3$ Mpc$^3$ in total, was 6 times higher than the average. They also evaluated the overdensity of LAEs which are typically 2 mag fainter than the LBGs in UV continuum to $R=25.5$ in the SSA22a field and found a similar value. \citet{key-hayashino} revisited the issue by using the deeper and wider-area data of LAEs. They confirmed the overdensity of $\sim 6$ around SSA22a in $10^3$ Mpc$^3$ and found the overdensity of $\sim 3$ even at the $\sim 10^5$ Mpc$^3$ volume.

We try to constrain the overdensity of such population of galaxies in the proto-cluster from the observed surface number density of DRGs in SSA22-M1. To obtain the number density of DRGs in a general field at $z\sim 3$, we adopt the luminosity function with $\phi^* = 6.14 \times 10^{-4}$ Mpc$^{-3}$ mag$^{-1}$, $M^* =-22.63$, and $\alpha = -0.46$ for DRGs at $2.7<z<3.3$ \citep{key-marchesini}. Our limiting $K$-band magnitude corresponds to $M_V=-20.8$ and the expected number density of DRGs above the luminosity is $5.9 \times 10^{-4}$ Mpc$^{-3}$. The volume of the proto-cluster sampled in SSA22-M1 ($6' \times 3.5'$ $\approx$ 11 Mpc $\times$ 7 Mpc in comoving scale) is $1.4 \times 10^3$ Mpc$^3$ if we assume the redshift range of $z=3.08-3.10$, which is similar to those studied for the excess density of LBGs or LAEs. Consequently, the expected average number of DRGs in the volume of SSA22-M1 with $\Delta z=0.02$ is obtained to be $\approx 0.8$.

We found at least three DRGs in vicinity of the two most luminous LABs (Table \ref{tab2}). As a result of the number density of DRGs, it is shown that approximately 5 DRGs at the redshift provides the similar overdensity as seen for LBGs or LAEs ($\approx 6 $ times the average). Although the discussion is limited by the large uncertainty due to the small number of the objects, an overdensity similar to those of LBGs/LAEs is allowed if the number of the foreground/background DRGs in the field is less than $\sim 80\%$ of that in  GOODS-N (29.5 DRGs are expected).

\section{Summary}

We presented the results of our deep near-infrared imaging observations of the $z=3.1$ proto-cluster in the SSA22a field taken by MOIRCS mounted on Subaru Telescope. We observed 21.7 arcmin${^2}$ field to the depths of $J=24.5$, $H=24.3$ and $K=23.9$ with 5$\sigma$. Our observed field covers the area where the surface number density of the LAEs is highest. 

We investigated whether massive and/or mature galaxies have already formed in the proto-cluster which is characterized by the overdensity of star-forming galaxies. We examined the distribution of the $K$-selected galaxies by using the simple color cut for DRGs as well as the photo-z selection. The surface number density of DRGs with $(J-K)_{AB} > 1.4$ in the field was 1.3 arcmin${^{-2}}$ at $K<23.4$. While it was not likely that the density of DRGs have the similar excess as seen for LBGs at the volume, we found significant evidence that more than a few galaxies with the stellar mass $M_* = 10^{9}-10^{11}$ M$_\odot$, exist in vicinity of, or might be associated with, the LABs. 

We also investigated all the $K_s$-band counterparts which have the consistent photometric redshift not only for the LABs but also for the other objects at $z=3.1$ in the field. We found that 88\% of the LABs have the plausible $K_s$-band candidates. The sum of the stellar mass of the galaxies possibly associated with LABs correlated with the luminosity and surface brightness of the LABs, which implied that the origin of Ly$\alpha$ emission must be closely related with the massive galaxy formation phenomena. The most luminous LABs, LAB1 and LAB2, had the $K$-selected counterparts with $M_* \sim10^{11}$ M$_\odot$. In addition, the marginal density excess of DRGs and the photo-z selected objects was found around the most luminous LABs, LAB1 and LAB2. 

Our results suggest that LABs are related to the formation of massive galaxies and a certain amount of evolved galaxies with $M_* \sim 10^{11} M_{\odot}$ are being formed in the central part of the high density region of star-forming galaxies at $z=3.1$. We are witnessing a stage when significant amount of the stars are being formed in the central part of the growing large-scale proto-cluster at $z=3.1$.

\vspace*{5mm}

We thank the staff of the Subaru Telescope for their assistance with the development and the observation of MOIRCS. This study is based on data collected at Subaru Telescope, which is operated by the National Astronomical Observatory of Japan. This research is supported in part by the Grant-in-Aid for Scientific Research 11554005 and 14340059 of the Ministry of Education, Science, Culture, and Sports in Japan. The Image Reduction and Analysis Facility (IRAF) used in this paper is distributed by the National Optical Astronomy Observatories, U.S.A., which are operated by the Association of Universities for Research in Astronomy, Inc., under cooperative agreement with the National Science Foundation.

\clearpage

\begin{table*}[htbp]
\begin{center}
\caption{The NIR photometric properties of $K$-selected objects in LABs \label{tab2}} 
\begin{tabular}{lcccccc} 
\hline 
\hline
LAB & Object & RA & Dec & $K$ & $J-K$ & $R-K$  \\
No. & ID \footnotemark[$*$] &  & &  & ($1''.1 \phi$) & ($2''.0 \phi$)  \\

\hline

LAB1 & C11 (\#5)      & 22:17:25.7 & 0:12:34.6 &       23.18 $\pm$   0.16 &     1.23 $\pm$   0.42 &     1.31 $\pm$   0.56  \\ 
     & \#1      & 22:17:26.1 & 0:12:46.7 &       22.23 $\pm$   0.10 &     1.08 $\pm$   0.28 &     4.08 $\pm$   0.51  \\ 
     & \#2      & 22:17:25.8 & 0:12:43.8 &       22.75 $\pm$   0.13 &     0.33 $\pm$   0.26 &     1.34 $\pm$   0.39  \\ 
     & \#3      & 22:17:26.0 & 0:12:36.4 &       23.07 $\pm$   0.15 &     1.48 $\pm$   0.45 &     2.77 $\pm$   0.51  \\ 
     & \#4      & 22:17:26.1 & 0:12:32.3 &       22.38 $\pm$   0.11 &     1.31 $\pm$   0.33 &     2.20 $\pm$   0.30  \\ 
\hline       
LAB2 & M14 \footnotemark[$\dagger$] (\#1)  & 22:17:39.1 & 0:13:30.6 &       23.30 $\pm$   0.16 &     1.60 $\pm$   0.47 &     2.60 $\pm$   0.64  \\ 
     & \#2      & 22:17:39.1 & 0:13:26.4 &       22.26 $\pm$   0.10 &     0.55 $\pm$   0.25 &     1.13 $\pm$   0.30  \\ 
     & \#3      & 22:17:38.9 & 0:13:24.0 &       19.98 $\pm$   0.03 &     0.50 $\pm$   0.07 &     1.85 $\pm$   0.03  \\ 
     & \#4      & 22:17:39.3 & 0:13:22.1 &       22.63 $\pm$   0.11 &     1.95 $\pm$   0.41 &     4.32 $\pm$   0.80  \\ 
\hline       
LAB7 & M4       & 22:17:41.0 & 0:11:27.8 &       23.78 $\pm$   0.19 &     0.23 $\pm$   0.34 &     0.90 $\pm$   1.02  \\ 
     & C6       & -          & -         &      - &     - &     - \\ 
\hline       
LAB8 & C15      & -          & -         &      - &     - &     - \\ 
\hline       
LAB16 &         & 22:17:24.9 & 0:11:17.5 &       23.29 $\pm$   0.16 &     0.91 $\pm$   0.36 &     1.78 $\pm$   0.59  \\ 
\hline       
LAB20 & C12     & 22:17:35.3 & 0:12:47.4 &       24.12 $\pm$   0.23 &     0.70 $\pm$   0.44 &     0.49 $\pm$   1.12  \\ 
\hline       
LAB30 & D3      & 22:17:32.5 & 0:11:32.9 &       22.88 $\pm$   0.13 &     0.73 $\pm$   0.31 &     1.08 $\pm$   0.41  \\ 
\hline       
LAB31 & C4      & 22:17:39.0 & 0:11:26.1  &       23.39 $\pm$   0.16 &     1.14 $\pm$   0.41 &     1.29 $\pm$   0.73  \\ 

\hline \hline
\multicolumn{4}{@{}l@{}}{\hbox to 0pt{\parbox{180mm}{\footnotesize
\footnotemark[$*$] ID of LBGs is referred from
 \citet{key-steidel03a}.
\par\noindent
\footnotemark[$\dagger$] The associated NIR object of M14 is located at $0''.9$ apart from
 the peak of the rest-frame UV source. 
}\hss}}
\end{tabular}
\end{center}
\end{table*}

\begin{table*}[htbp]
\begin{center}
\caption{The photometric redshift and stellar mass of NIR counterparts \label{tab3}} 
\begin{tabular}{lccccccc} 
\hline \hline
LAB & Object & $z_{\rm spec}$ \footnotemark[$\dagger$] & $z_{\rm photo}$ &  Age \footnotemark[$\ddagger$]
 & E(B-V) \footnotemark[$\ddagger$]  & Mv \footnotemark[$\ddagger$]  & Stellar Mass \footnotemark[$\ddagger$]   \\
  & ID \footnotemark[$*$]    &  &   & (Gyr) & (mag)  & (mag) & ($10^{10} M_{\odot}$) \\ 
\hline

LAB1&   C11     &   3.109 &   2.83    &     1.80 &     0.08 &   -22.32 &      2.5  $_{-     1.5}  ^{+     2.4}$\\ 
    &   \#3     &   -    &   2.85    &     1.80 &     0.20 &   -22.59 &      7.2  $_{-     2.5}  ^{+     4.3}$\\ 
    &   \#4     &   -    &   2.61    &     1.61 &     0.18 &   -23.19 &     10.9  $_{-     5.0}  ^{+     4.8}$\\ 
\hline
LAB2&   M14  \footnotemark[$\S$]  &  3.091 &   3.19    &     1.80 &     0.22 &   -22.64 &      8.1  $_{-     6.9}  ^{+     4.8}$\\ 
    &  \#4      &   -    &   2.61   &     0.20 &     0.40 &   -23.05 &     10.5  $_{-     6.3}  ^{+    32.6}$\\ 
\hline
LAB7&   M4      &   3.093 &   2.70    &     0.57 &     0.06 &   -21.75 &      0.7  $_{-     0.7}  ^{+     2.8}$\\ 
\hline
LAB16&  -       &   -    &   2.70    &     0.06 &     0.20 &   -22.50 &      1.6  $_{-     1.2}  ^{+     3.6}$\\ 
\hline
LAB20&  C12     &   3.118 &   2.85    &     0.57 &     0.00 &   -21.43 &      0.4  $_{-     0.3}  ^{+     1.1}$\\ 
\hline
LAB30&  D3      &   3.086 &   3.31    &     1.61 &     0.08 &   -22.74 &      3.4  $_{-     2.1}  ^{+     1.1}$\\ 
\hline
LAB31&  C4      &   3.076 &   3.19    &     1.80 &     0.00 &   -22.04 &      2.0  $_{-     1.4}  ^{+     1.5}$\\

\hline \hline
\multicolumn{4}{@{}l@{}}{\hbox to 0pt{\parbox{180mm}{\footnotesize
\footnotemark[$*$] ID of LBGs is referred from
 \citet{key-steidel03a}.
\par\noindent
\footnotemark[$\dagger$] The redshifts of LBGs are referred from
 \citet{key-steidel03a}.
\par\noindent
\footnotemark[$\ddagger$] The redshifts are assumed to be $z=3.1$ when the
 SEDs are calculated.
\par\noindent
\footnotemark[$\S$] The associated NIR object of M14 is located at $0''.9$ apart from the peak of the rest-frame UV source. 
}\hss}}
\end{tabular}
\end{center}
\end{table*}

\onecolumn

\end{document}